\documentstyle[twocolumn,prb,aps,epsfig]{revtex}

\begin{document}
\title{ {\it Ab initio} investigation of VOSeO$_3$, 
a spin gap system with coupled spin dimers}

\author{Roser Valent\'\i$^1$, T. Saha-Dasgupta$^2$ and F. Mila$^3$ } 

\address{$^1$Fakult\"at 7, Theoretische Physik,
 University of the Saarland,
66041 Saarbr\"ucken, Germany.}

\address{$^2$S.N. Bose National Centre for Basic Sciences, 
JD Block, Sector 3,
 Salt Lake City, Kolkata 700098, India.}

\address{$^3$Institut de Physique Th\'eorique,
 Universit\'e de Lausanne,
CH-1015 Lausanne, Switzerland.}

\date{\today}
\maketitle


\begin{abstract}
Motivated by an early experimental study of VOSeO$_3$, 
which suggested that it is a quasi-2D system
of weakly coupled spin dimers with a small spin gap, 
we have investigated the electronic structure of this material
via density-functional calculations. These {\it ab initio} 
results indicate that the system is better thought of as an 
alternating spin-1/2 chain with moderate interchain
interactions, an analog of (VO)$_2$P$_2$O$_7$. 
The potential interest of this system for studies in high magnetic
field given the presumably small value of the spin gap is emphasized.
\end{abstract}


\vspace{1cm}

The search for low-dimensional s=$\frac{1}{2}$ materials with a 
singlet-triplet gap in the spin excitation spectrum has been a very
active field of research since the discovery of high T$_c$ 
superconducting cuprates. The original motivation came from the possible
connection between spin-gap and superconductivity, an issue still
intensely debated in the context of cuprates. It became clear
however with the extensive investigation of spin ladders
that even purely insulating magnets with a spin gap can exhibit
a very rich physics\cite{dagotto,chaboussant}. 
The basic idea is quite simple\cite{mila}: When the gap
of a system consisting of weakly coupled dimers is closed by a 
strong enough magnetic field, some dimers are promoted to triplets.
These triplets behave as a gas of quantum particles, and they
can undergo different kinds of phase transitions as a 
function of magnetic field or temperature depending on 
dimensionality, topology, etc... 
To reach this regime however, one needs systems with 
moderate gaps - up to 20 K or so. This is the typical 
range of magnetic coupling in organic materials, and 
the first investigations have been accordingly 
carried out on organic systems. These systems are tricky however, 
and even the most extensively studied organic ladder, 
Cu$_2$(C$_5$H$_{12}$N$_2$)$_2$Cl$_4$,
is still a subject of debate and 
controversies\cite{chaboussant,broholm}. The difficulties
with organic systems are two-fold: the exchange paths are not
directly evident on the basis of the structure, and it is in most cases
impossible to grow large single crystals. 

These problems are usually much easier to solve in inorganic
transition-metal compounds. The main difficulty there comes 
from the order of magnitude of the exchange 
integrals - typically several hundreds of degrees Kelvin -
but if the geometry is such that the exchange integrals are 
reduced substantially in the spirit
of the Goodenough-Kanamori-Anderson 
rules\cite{GKA}, then the low-temperature properties are
usually easier to reach. Several qualitatively different
classes of behaviour have been 
identified so far in this family, among which are incommensurate spin-spin
correlations
in the spin-Peierls high-field phase of CuGeO$_3$\cite{boucher},
magnetic
field-induced 
long-range order in TlCuCl$_3$ and KCuCl$_3$\cite{cavadini,SD_02},
 and magnetization 
plateaux in SrCu$_2$(BO$_3$)$_2$\cite{kageyama} due to
magnetic field-induced localization of spin-triplets. The behaviours
reported so far do not exhaust the possibilities though. 

In that respect, the vanadyl system VOSeO$_3$ is potentially
very interesting. This compound was first investigated by Trombe
{\it et al.}\cite{Trombe_87}, who synthetized the system, determined
its structure, and measured the temperature dependence of the 
susceptibility. They concluded that the system
consists of weakly coupled dimers, and that the residual couplings
are presumably of two-dimensional character. 
Besides, although the susceptibility data reported by 
Trombe {\it et al.} do not allow the determination of the 
spin gap because of a large Curie tail, the location of 
the maximum of the susceptibility suggests that the gap is 
probably in the appropriate range to be closed by 
a magnetic field.
However, the details of the dominant interactions, their origin 
as well as the role of the active orbitals at the Fermi surface 
could not be quantitatively discussed in that work, and the conclusions
are only preliminary. 

In order to check the validity of this analysis, and in particular
to determine whether the couplings are indeed essentially 
two-dimensional, we have performed an {\it ab initio} 
calculation of the electronic properties of this system, followed
by a tight-binding-downfolding analysis in order to define
the important hopping parameters in this compound. 
As we shall see, the emerging picture is  somewhat different,
although not less interesting, than that guessed by 
Trombe {\it et al.}\cite{Trombe_87}.

\begin{figure}[t]

\vspace*{0.0cm}

(a)
\hspace*{0.15cm}\centerline{\hspace*{0.3cm}
\epsfig{file=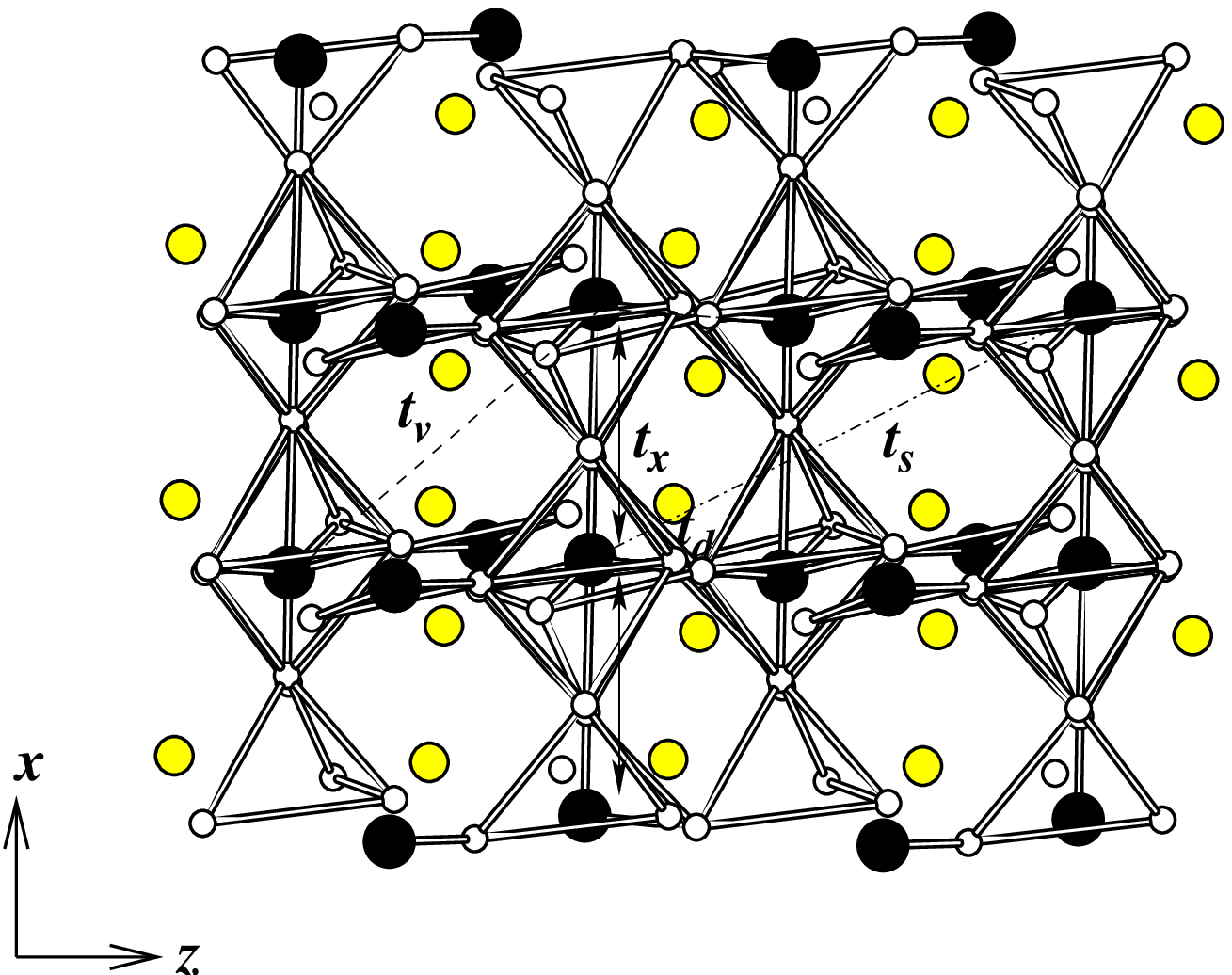,width=0.5\textwidth}
           }

(b)
\centerline{\hspace*{0.15cm}
\epsfig{file=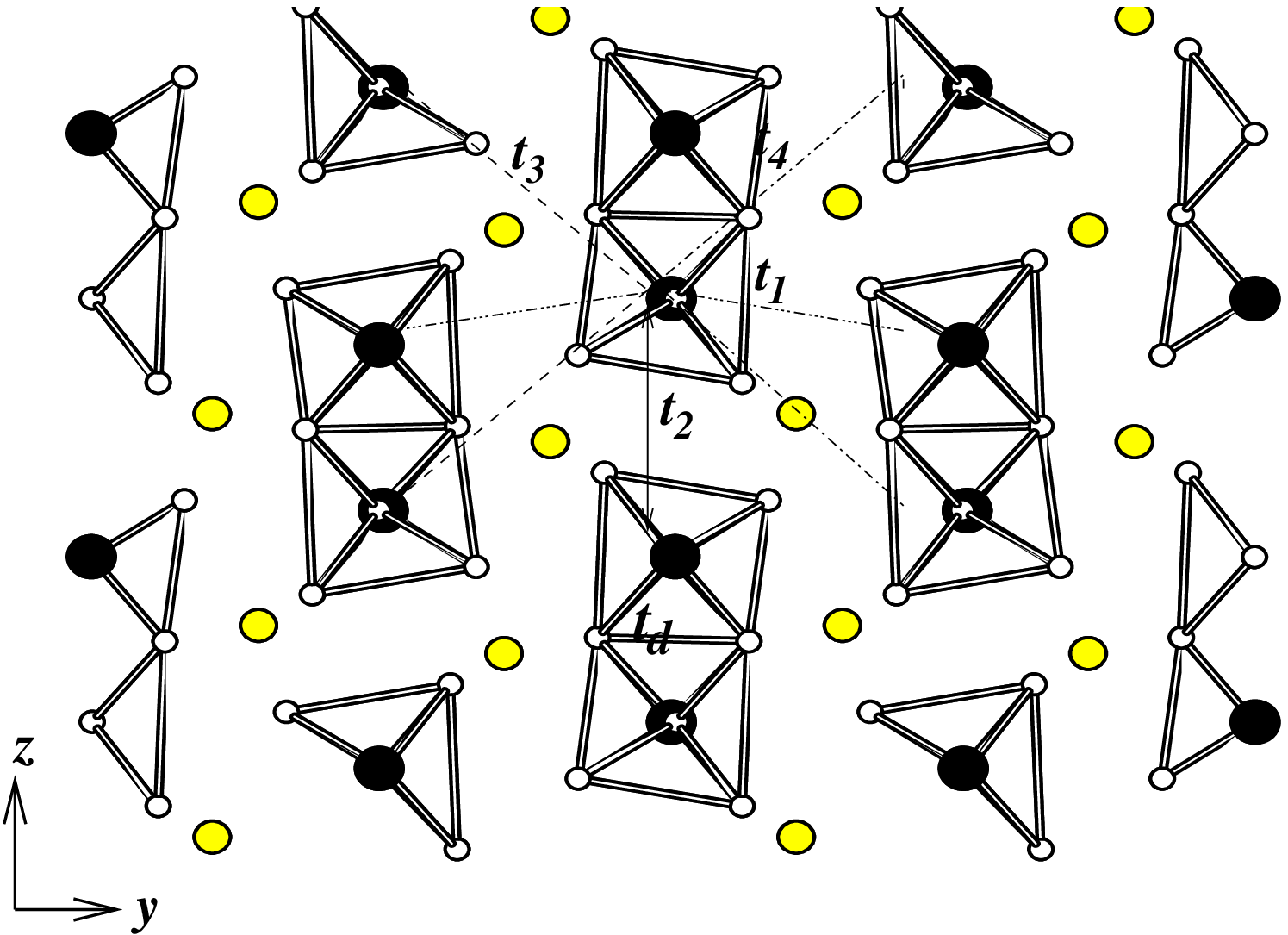,width=0.5\textwidth,angle=0}
 }
\vspace*{0.1cm}
\caption{\label{structure}
Crystal structure of VOSeO$_3$ projected in the (a) $zx$ plane and
(b) $yz$ plane.  The large black circles are V$^{4+}$ in a square
 pyramid environment of oxygens (small white circles). Se atoms are
 represented by grey balls. Shown in this figure are also the
 relevant hoppings in this material. }
\end{figure}

In order to understand the electronic behavior of this system, 
an examination of the crystal structure is  essential.
VOSeO$_3$ crystallizes in the 
monoclinic space-group P2$_1$/c with lattice parameters  $a$ = 4.0168 $\AA$,
$b$ = 9.788 $\AA$, $c$= 8.001 $\AA$, $\beta$ = 99.42$^o$, and it
contains four formula units per
primitive unit cell\cite{Trombe_87}.  The vanadium ions V$^{4+}$
 form square pyramids
with the neighboring oxygens. Two adjacent up and down square pyramids
share an edge and build dimeric [V$_2$O$_8$]$^{8-}$ units. These units 
form chains along the $x$ direction (see Fig.\ \ref{structure} (a))
 and are linked through Se atoms in the
$yz$ plane as shown in Fig.\ \ref{structure} (b). 
The vanadium atom in the square pyramid is shifted towards the apex
 oxygen and forms a short vanadium-oxygen bond ($d=1.611 \AA$) characteristic
 of a vanadyl ion VO$^{2+}$. The distance between two V$^{4+}$ ions
 within the dimeric unit is of $d=3.175 \AA$ while the distance between two
 dimeric units along $x$ is $d= 4.015 \AA$ and along $z$ is $d=4.841 \AA$.

  We have carried out a first-principles  study
 based on the density-functional
 theory (DFT) 
 in order to derive  the electronic properties of VOSeO$_3$.
 We have used the generalized gradient approximation (GGA)\cite{Perdew_96}
 in order to include the non-local effects within the gradient approximation
  to go beyond the local density approximation (LDA).
 Calculations have
  been performed  within the
 framework of both the full-potential
 linearized augmented plane wave (LAPW) method based on
  WIEN97\cite{WIEN97} code and the
 linearized muffin tin orbital (LMTO)\cite{Andersen_75} method based
on the Stuttgart TBLMTO-47 code. The results obtained by both methods
 are in agreement with each other.

\begin{figure}[t]

\vspace*{0.0cm}

\centerline{\hspace*{0.6cm}
\epsfig{file=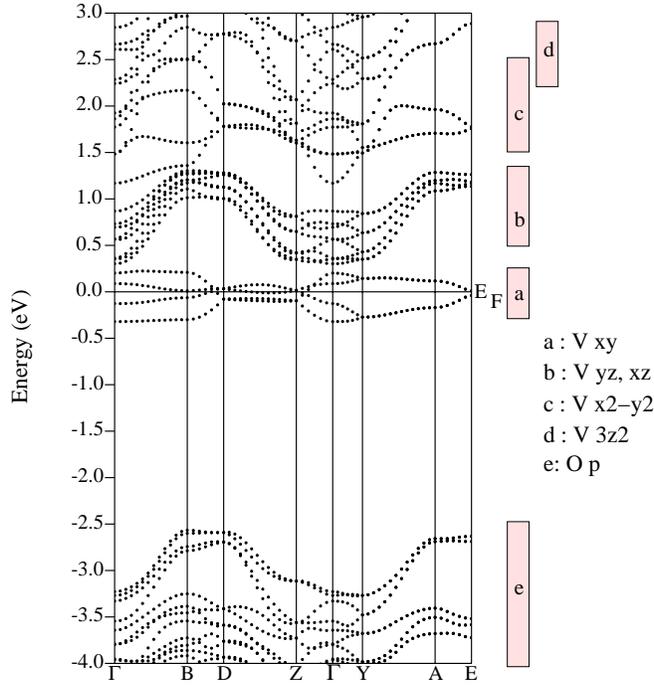,width=0.5\textwidth}
          }
\vspace*{0.3cm}
\caption{\label{bands}
 Band-structure for VOSe$_3$ along the path $\Gamma$-B-D-Z-$\Gamma$-Y-E.
 The boxes indicate the band character in the local coordinate system.}

\end{figure}

In Fig.\ \ref{bands} we show a plot of the band structure along the symmetry
 path in the Brillouin zone $\Gamma$ =
 (0, 0, 0), B=(-$\pi$,0,0), D=(-$\pi$,0,$\pi$), Z=(0, 0, $\pi$), $\Gamma$,
 Y=(0,$\pi$,0), A=(-$\pi$, $\pi$, 0), E=(-$\pi$, $\pi$, $\pi$).  
      This system shows four narrow bands 
(four since the unit cell has four vanadium atoms) at the Fermi level
 with a bandwith of about 0.5 eV.  These bands are of vanadium 3d$_{xy}$ character
 (in the local frame of reference with the z axis pointing along the axis
 connecting the vanadium ion and the apical oxygen of the square pyramid)
 with small mixing of oxygen $p$ characters.
   They are separated by a gap of about 2 eV  
 from the lower valence bands and  a gap of 0.1 eV from the higher-unoccupied
 conduction bands. The system is half-filled and the insulator behavior
 observed in this system should be explained by the effect of electron
 correlation which is not taken fully into account in the  LDA or GGA
 calculations. 
   Note that the dispersion along the $x$ axis 
 (chain direction) is small while in the $yz$ plane the dispersion is of about
 200 meV. This behavior already indicates that the important
 interactions
in this system will be in the $yz$ plane. 
 We shall analyze this point in more detail
 in terms of an effective hamiltonian with parameterized hopping integrals.

\begin{figure}[t]

\vspace*{0.1cm}

\centerline{\hspace*{1.5cm}
\epsfig{file=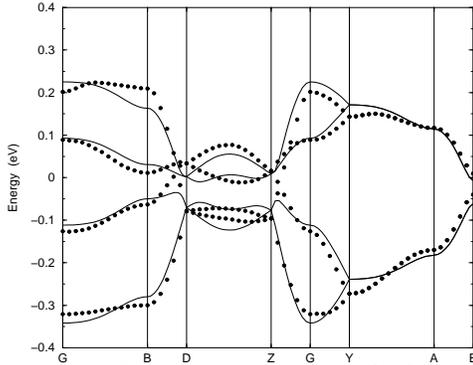,width=0.35\textwidth,angle=-90}
           }
\caption{\label{TB_model}
Comparison of the {\it downfolded}-tight-binding bands (solid lines)
with the DFT bands (dotted lines). The tight-binding parameters (see
 Fig.\ \protect\ref{structure}) are (in eV) t$_d$= 0.083, $t_2$= 0.079,
 $t_1$= 0.012, $t_3$= 0.005, t$_4$= 0.040, $t_v$= 0.040, $t_s$=0.016,
 $e_0$=-0.034, $t_x$= 0.0.
}
\end{figure}

\begin{figure}[t]


\centerline{\hspace*{0.0cm}
\epsfig{file=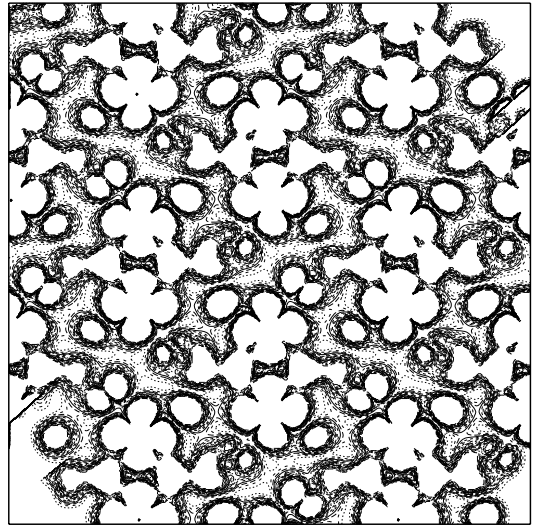,width=0.55\textwidth}
           }
\caption{\label{ED}
 Projection on the $zy$ plane of the electron density for
 bands close to the Fermi level (compare with Fig.\
 \protect\ref{structure}).
 Note the Vd$_{xy}$-Op-Vd$_{xy}$
 pathes contributing to the intradimer coupling in VOSeO$_3$.
}
\end{figure}

  We have employed LMTO-based 
downfolding\cite{newlmto} and tight-binding analysis on the {\it ab initio}
 results to obtain an effective few-orbital tight-binding description of
this material.  The downfolding method consists in deriving a few-orbital effective 
Hamiltonian from the full  LDA or GGA  Hamiltonian by downfolding the inactive 
orbitals in the tails of the active orbitals kept in the basis chosen
to describe the low-energy physics. This process results in 
 renormalized effective interactions
between the active orbitals,  V d$_{xy}$ in the present case.
  By Fourier transforming the downfolded
hamiltonian $H_k \rightarrow H_R$,
\begin{eqnarray}
H_R = -\sum_{i,j}t_{ij}(c_j^{\dagger}c_i + c_i^{\dagger}c_j)
\end{eqnarray}
one can  then  extract the effective hopping matrix elements $t_{ij}$
between the vanadium ions.  The comparison between
downfolded-tight-binding bands and the DFT bands are shown in Fig.\
\ref{TB_model}.
 The band dispersion of the four-band complex close
to the Fermi level can be reproduced well by considering a few short-ranged 
 hopping parameters (see Fig.\ \ref{structure}): $t_d$ is the intradimer hopping
 between two edge-sharing V$^{4+}$O$_5$ pyramids and
 it is expected to give the largest contribution since it 
corresponds to  a superexchange
 path V-O-V  ($\alpha$ $\sim$ 103$^o$)\cite{Beltran_89}. Its 
 value is essentially related  to the width of the four-band set.
 In Fig.\ \ref{ED} we show the projection of the electron density on
 the $zy$ plane  (compare with Fig.\ \ref{structure}(b) ),
where the contribution 
 of the Vd$_{xy}$-Op-Vd$_{xy}$ path to the intradimer coupling can
 be clearly identified.

 The next important hopping integral is $t_2$ which des\-cribes the interdimer
 coupling along the $z$ direction. Other hopping parameters in the 
 $yz$ plane, $t_4$, $t_1$, $t_3$ are smaller but non-negligible. They are
 responsible for the band-splitting along $\Gamma D$ and dispersion
 along $ZY$. Note that the pathes described by 
$t_3$ and $t_4$ are not equivalent as
 can be observed in the electron density plot of Fig.\ \ref{ED}
 (compare
 with Fig.\ \ref{structure}).
  The hopping integral along the chain direction $x$, $t_x$
 proves to be very small, which indicates that this interaction path is
 negligible. 
 This is to be expected from the fact that the active d$_{xy}$ orbital defined
 in the rotated local frame of reference with the $z$ axis turned along
 the vanadium -apical oxygen, has very weak interaction with the apical
 oxygen $p$
 orbital.
 Still there are two more parameters $t_v$ and $t_s$ which are not
 to be neglected since they are responsible for the band splitting along
 the path $DZ$ and dispersion along $YA$.
  This system shows a very similar behavior to CsV$_2$O$_5$\cite{Valenti_02}
 as well as (VO)$_2$P$_2$O$_7$\cite{Garrett_97} though in those cases
 important pathes of interaction where  provided through
V$^{5+}$O$_4$ and P$^{5+}$O$_4$ tetrahedra groups respectively,
which are not present in   VOSeO$_3$.  Instead, Se$^{4+}$O$_3$ trigonal 
pyramid groups contribute decisively to the interdimer interaction
pathes in VOSeO$_3$. Those groups can be considered as playing
the equivalent role to the tetrahedra groups in the above mentioned 
compounds.

A detailed comparison between the hopping parameters in VOSeO$_3$
and CsV$_2$O$_5$\cite{Valenti_02} confirms the similitude of
their behavior 
(compare $t_d$, $t_2$, $t_4$ with $t_1$, $t_3$, $t_5$ in
Ref.\ \cite{Valenti_02}). There is a quantitative  distinction
though between both systems which is related to the band-splitting in
VOSeO$_3$ along the path $DZ$ described by the hopping 
parameters $t_v$ and $t_s$ which is not present in  CsV$_2$O$_5$. 
  
The analysis of the effective tight-binding model for VOSeO$_3$
leads us to conclude that this system shows a coupled spin
dimer behaviour with important interdimer coupling along the $z$
direction and moderate to negligible couplings along $y$ and $x$.
This picture 
provides a quantitative description of the important interactions
in VOSeO$_3$  as opposed to that of
Trombe {\it et al.}\cite{Trombe_87} on the basis of their susceptibility
data.

 An estimate of the exchange integral related to the
 dominating interdimer interaction parameter $t_d$ can be obtained
by using the relation J $\sim$ $4 t_d^2/U$ where $U$ is the effective
 onsite Coulomb repulsion on the vanadium site. Such an estimate
 is valid since the path described by $t_d$ corresponds to
 a V-O-V superexchange path.  Values of U $\sim$ 4-5 eV have been
 proposed for other vanadyl systems\cite{Rosner_02}.
Assuming that this value is similar for VOSeO$_3$ we get a J $\sim$
  5 meV
 $\sim$ 55 K
which is of the same order of magnitude as
the J value obtained by Trombe {\it et al.} (J $\sim$ 30K) from the analysis of
their susceptibility data.

  We hope that the present results will motivate further
experimental efforts to understand the properties of this
system. It would be particulartly interesting to investigate its
properties under high magnetic field since its topology
is {\it a priori} different from that of other materials in which 
the spin gap could be closed.
 
We acknowledge useful discussions with P. Millet and the support
of the Swiss National Fund and the Deutsche Forschungsgemeinschaft.




\vspace*{0.2cm}

\

\end{document}